\documentclass[apj]{emulateapj}

\makeatletter
\usepackage{amsmath}
\makeatother

\begin{document}

\title{Search for Diffuse Astrophysical Neutrino Flux Using Ultra--High-Energy Upward-going Muons in Super-Kamiokande I}
\shorttitle{Astrophysical Neutrinos with UHE Muons in SK}
\shortauthors{Swanson et al.}

\def\icrr{1}
\def\ncen{2}
\def\bu{3}
\def\bnl{4}
\def\uci{5}
\def\csu{6}
\def\cnu{7}
\def\duke{8}
\def\gmu{9}
\def\gifu{10}
\def\uh{11}
\def\ui{12}
\def\kek{13}
\def\kobe{14}
\def\kyoto{15}
\def\lanl{16}
\def\lsu{17}
\def\umd{18}
\def\mit{19}
\def\duluth{20}
\def\miyagi{21}
\def\nagoyaste{22}
\def\suny{23}
\def\niigata{24}
\def\okayama{25}
\def\osaka{26}
\def\seoul{27}
\def\shizuoka{28}
\def\shizuokafukushi{29}
\def\skku{30}
\def\tohoku{31}
\def\tokai{32}
\def\tit{33}
\def\tokyo{34}
\def\warsaw{35}
\def\uw{36}
\def\corresponding{*}
\def\pennnow{$\dagger$}
\def\kamiokanow{$\ddagger$}
\def\icrrnow{\S}
\def\triumfnow{\P}
\def\stanfordnow{$||$}
\def\columbianow{**}
%
\newcommand{\tempaltaffiltext}[1]{$^{#1}$}
\author{
M.E.C.~Swanson\altaffilmark{\mit,\corresponding},
K.~Abe\altaffilmark{\icrr}, 
J.~Hosaka\altaffilmark{\icrr},
T.~Iida\altaffilmark{\icrr},  
K.~Ishihara\altaffilmark{\icrr},
J.~Kameda\altaffilmark{\icrr},
Y.~Koshio\altaffilmark{\icrr},
A.~Minamino\altaffilmark{\icrr},
C.~Mitsuda\altaffilmark{\icrr},
M.~Miura\altaffilmark{\icrr}
% ...
\\
(The Super-Kamiokande Collaboration)\\
\footnotesize
\rm
\tempaltaffiltext{\icrr}{Kamioka Observatory, Institute for Cosmic Ray Research, 
University of Tokyo, Kamioka, Gifu, 506-1205, Japan }\\
\tempaltaffiltext{\ncen}{Research Center for Cosmic Neutrinos, Institute for Cosmic 
Ray Research, University of Tokyo, Kashiwa, Chiba 277-8582, Japan }\\
\tempaltaffiltext{\bu}{Department of Physics, Boston University, Boston, MA 02215, 
USA }\\
\tempaltaffiltext{\bnl}{Physics Department, Brookhaven National Laboratory, Upton, 
NY 11973, USA }\\
\tempaltaffiltext{\uci}{Department of Physics and Astronomy, University of 
California, Irvine, Irvine, CA 92697-4575, USA }\\
\tempaltaffiltext{\csu}{Department of Physics, California State University, 
Dominguez Hills, Carson, CA 90747, USA }\\
\tempaltaffiltext{\cnu}{Department of Physics, Chonnam National University, Kwangju 
500-757, Korea }\\
\tempaltaffiltext{\duke}{Department of Physics,Duke University, Durham, NC 27708, USA }\\
\tempaltaffiltext{\gmu}{Department of Physics, George Mason University, Fairfax, VA 
22030, USA }\\
\tempaltaffiltext{\gifu}{Department of Physics, Gifu University, Gifu, Gifu 
501-1193, Japan }\\
\tempaltaffiltext{\uh}{Department of Physics and Astronomy, University of Hawaii, 
Honolulu, HI 96822, USA }\\
\tempaltaffiltext{\ui}{Department of Physics, Indiana University, Bloomington,
 IN 47405-7105, USA }\\ 
\tempaltaffiltext{\kek}{High Energy Accelerator Research Organization (KEK), 
Tsukuba, Ibaraki 305-0801, Japan }\\
\tempaltaffiltext{\kobe}{Department of Physics, Kobe University, Kobe, Hyogo 
657-8501, Japan }\\
\tempaltaffiltext{\kyoto}{Department of Physics, Kyoto University, Kyoto 606-8502, 
Japan }\\
\tempaltaffiltext{\lanl}{Physics Division, P-23, Los Alamos National Laboratory, Los 
Alamos, NM 87544, USA }\\
\tempaltaffiltext{\lsu}{Department of Physics and Astronomy, Louisiana State 
University, Baton Rouge, LA 70803, USA }\\
\tempaltaffiltext{\umd}{Department of Physics, University of Maryland, College Park, 
MD 20742, USA }\\
\tempaltaffiltext{\mit}{Department of Physics, Massachusetts Institute of 
Technology, Cambridge, MA 02139, USA }\\
\tempaltaffiltext{\duluth}{Department of Physics, University of Minnesota, 
Duluth, MN 55812-2496, USA }\\
\tempaltaffiltext{\miyagi}{Department of Physics, Miyagi University of Education, Sendai,Miyagi 980-0845, Japan }\\
\tempaltaffiltext{\nagoyaste}{Solar-Terrestrial Environment Laboratory, Nagoya University, Nagoya, Aichi 464-8601, Japan }\\
\tempaltaffiltext{\suny}{Department of Physics and Astronomy, State University of 
New York, Stony Brook, NY 11794-3800, USA }\\
\tempaltaffiltext{\niigata}{Department of Physics, Niigata University, Niigata, 
Niigata 950-2181, Japan }\\
\tempaltaffiltext{\okayama}{Department of Physics, Okayama University, Okayama, Okayama 700-8530, Japan }\\
\tempaltaffiltext{\osaka}{Department of Physics, Osaka University, Toyonaka, Osaka 
560-0043, Japan }\\
\tempaltaffiltext{\seoul}{Department of Physics, Seoul National University, Seoul 
151-742, Korea }\\
\tempaltaffiltext{\shizuoka}{Department of Systems Engineering, Shizuoka University, 
Hamamatsu, Shizuoka 432-8561, Japan }\\
\tempaltaffiltext{\shizuokafukushi}{Department of Informatics in
Social Welfare, Shizuoka University of Welfare, Yaizu, Shizuoka 425-8611, Japan }\\
\tempaltaffiltext{\skku}{Department of Physics, Sungkyunkwan University, Suwon 
440-746, Korea }\\
\tempaltaffiltext{\tohoku}{Research Center for Neutrino Science, Tohoku University, 
Sendai, Miyagi 980-8578, Japan }\\
\tempaltaffiltext{\tokai}{Department of Physics, Tokai University, Hiratsuka, 
Kanagawa 259-1292, Japan }\\
\tempaltaffiltext{\tit}{Department of Physics, Tokyo Institute for Technology, 
Meguro, Tokyo 152-8551, Japan }\\
\tempaltaffiltext{\tokyo}{The University of Tokyo, Tokyo 113-0033, Japan }\\
\tempaltaffiltext{\warsaw}{Institute of Experimental Physics, Warsaw University, 
00-681 Warsaw, Poland }\\
\tempaltaffiltext{\uw}{Department of Physics, University of Washington, Seattle, WA 
98195-1560, USA }
}
\altaffiltext{\corresponding}{Electronic address: molly@space.mit.edu}
\altaffiltext{\pennnow}{Present address: Center for Gravitational Wave Physics, Pennsylvania State University, University Park, PA 16802, USA}
\altaffiltext{\kamiokanow}{Present address: Kamioka Observatory, Institute for Cosmic Ray Research, University of Tokyo, Kamioka, Gifu, 506-1205, Japan}
\altaffiltext{\icrrnow}{Present address: Institute for Cosmic Ray Research, University of Tokyo, Kashiwa, Chiba 277-8582, Japan}
\altaffiltext{\triumfnow}{Present address: TRIUMF, Vancouver, British Columbia V6T 2A3, Canada}
\altaffiltext{\stanfordnow}{Present address: Department of Physics, Stanford University, Stanford, CA 94305, USA}
\altaffiltext{\columbianow}{Present address: Department of Physics, Columbia University, New York, NY 10027, USA}

\slugcomment{ApJ in press, 652:206-215, 2006 November 20}

\begin{abstract}
Many astrophysical models predict a diffuse flux of high-energy neutrinos
from active galactic nuclei and other extra-galactic sources. At muon
energies above 1 TeV, the upward-going muon flux induced by neutrinos from
active galactic nuclei is expected to exceed the flux due to atmospheric
neutrinos. We have performed a search for this astrophysical neutrino
flux by looking for upward-going muons in the highest energy data sample 
from the Super-Kamiokande detector using 1679.6 live days of data.
We found one extremely high energy upward-going muon event, 
compared with an expected atmospheric neutrino background of 
$0.46 \pm 0.23$ events. Using this result, we set an upper limit on the diffuse flux 
of upward-going muons due to neutrinos from astrophysical sources in the muon energy 
range 3.16--100~TeV.
\end{abstract}


\maketitle

\section{Introduction}

\label{sec:Introduction}

The GeV-PeV energy range is unexplored territory for neutrino astronomy --- 
observations of neutrinos at these energies will open a new window
on the high energy universe.
A wide variety of astrophysical phenomena are expected to
produce extremely high energy neutrinos, ranging from active galactic
nuclei (AGNs) and gamma-ray bursts (GRBs; 
\citealt{2002RPPh...65.1025H,1995PhR...258..173G}) 
to more exotic sources such
as dark matter annihilation or decays of topological 
defects \citep{2004IJMPA..19..317S}.

The flux of neutrinos at such high energies is quite small; therefore,
large-scale detectors are required. One effective technique for observing
high-energy neutrinos with an underground detector is to look for
muons produced by $\nu_{\mu}$ or $\bar{\nu}_{\mu}$ interacting in
the surrounding rock. (Throughout this paper, ``muons'' will refer
to both $\mu^{+}$ and ${\mu^{-}}$.) The muon range in rock increases
with muon energy, which expands the effective interaction volume for
high-energy events. Downward-going neutrino-induced muons cannot be distinguished
from the much larger flux of downward cosmic-ray muons, but since cosmic 
ray muons cannot travel through the entire Earth, upward-going muons are 
almost always neutrino-induced. Thus, upward-going muons provide a suitable 
high-energy neutrino signal.

At muon energies above $1-10{\rm {\; TeV}}$, the upward-going muon
flux due to neutrinos from AGNs is expected to exceed the flux due
to atmospheric neutrinos~\citep{Stecker:1995th,Mannheim:1998wp}.
This cosmic
neutrino flux could be detected either by searching for point sources
of high-energy neutrinos or by detecting a diffuse, isotropic flux
of neutrinos from unresolved astrophysical sources. A diffuse cosmic
neutrino flux would be observed as an excess to the expected atmospheric
neutrino flux at high energies. In this analysis, we focus on searching
for a diffuse flux of upward-going muons due to neutrinos from astrophysical
sources using the highest energy data sample in Super-Kamiokande (Super-K).  This study
complements other Super-K searches for astrophysical point sources
of high energy neutrinos that use data over a larger
energy range \citep{2006ApJ...652..198A}.
In this paper we describe a search for evidence of a high energy astrophysical
neutrino flux in Super-K's highest energy upward-going muon sample. 
In \S~\ref{sec:Detector} we describe the 
Super-Kamiokande detector, and in \S~\ref{sec:Event-Selection} we give the 
details of how we selected candidate events from Super-K's ultra--high-energy sample. 
We evaluate 
our selection process with Monte Carlo 
in \S~\ref{sec:High-Energy-Isotropic-MC} 
and calculate the observed upward-going muon flux in \S~\ref{sec:Flux-Calculation}. 
Sections \ref{sec:Expected-Atmospheric-Background} and \ref{sec:Analytical-Estimate}
discuss the background due to the atmospheric neutrino flux. Based on the 
results, we set an upper limit in \S~\ref{sec:Upper-Limit} and conclude in 
\S~\ref{sec:Conclusions}. Any necessary estimates and approximations have been
 made so that they lead to a conservative result for this upper
limit.

\section{The Super-Kamiokande Detector}

\label{sec:Detector}

The Super-K detector is a cylindrical 50 kiloton
water Cerenkov detector, located in the Kamioka-Mozumi mine in Japan.
It is 41.4 m tall and 39.3 m in diameter.
The detector was constructed under the peak of Mount Ikenoyama, which
provides an average rock overburden
of 1000~m (2700~m water equivalent). Its geodetic location is at 
36.4{$^\circ$}~north, 137.3{$^\circ$}~east, and altitude 370~m. 

Super-K consists of 
two concentric, optically separated detectors. 
Until 2001 July the inner detector (ID) was 
instrumented with 11,146 inward-facing 50 cm diameter photomultiplier tubes 
(PMTs). The outer detector (OD) is a cylindrical shell of water surrounding
the ID and is instrumented with 1885 outward-facing 20~cm diameter
PMTs. Between the ID and the OD, there is a 50~cm thick shell.
Photons coming from this
region will not be detected by either the OD or the ID, so we refer
to it as the insensitive region. 

More details about the detector can
be found in \citet{2003NIMPA.501..418T}.
The data sample used in this analysis
was taken from 1996 April to 2001 July, corresponding to 1679.6 days
of detector livetime. This data run is referred to as SK-I.

Super-K is primarily designed to detect lower energy neutrinos from
the Sun, the atmosphere, and particle accelerators 
but can potentially
detect the extremely high energy neutrinos expected from astrophysical
sources as well. This paper focuses on the
events at the highest energy end of Super-K's detection range.

\section{Event Selection}
\label{sec:Event-Selection}
The ultra--high-energy sample in SK-I consists of events that
deposit $\ge1.75\times10^{6}$ photoelectrons (pe) in the ID.
In the low-energy regime, on average about 9 pe are recorded by the ID PMTs
for each MeV of energy deposited in the tank;
the electronics for the ID PMTs saturate at about 300 pe. Thus an event
with $\ge1.75\times10^{6}$ pe in the ID corresponds to a minimum energy
deposition of approximately 200 GeV, but the actual energy deposition
could be much higher, since the saturation effect prevents
all of the produced pe from being recorded.

At high energies, muons have some probability to lose energy through radiative
processes such as bremsstrahlung, resulting in an electromagnetic shower
that deposits large quantities of pe in the detector. For comparison, a muon
that traverses the maximum path length through the ID (50 m) but does not
produce any electromagnetic showers will deposit approximately 11 GeV via 
ionization energy loss, corresponding to  $\sim 10^{5}$ pe deposited in the 
ID. Thus a high-pe
cutoff offers a means of selecting high energy events. 

At the high-pe threshold of $\ge1.75\times10^{6}$ pe, the high level of 
saturation in the ID PMT electronics 
can cause Super-K's precision
muon fitting algorithms to fail. Therefore, these extremely energetic events
are not included in other studies of upward-going muons in SK-I 
\citep{2004PhRvD..70h3523D,1999PhRvL..82.2644F,2005PhRvD..71k2005A}.
In this study we analyzed this ultra--high-energy data sample separately 
using a different fitting method based on information
from the OD.

\subsection{Outer Detector Linear Fit}

\label{sub:Linear-Fit-Method}

\begin{figure*}
\plottwo{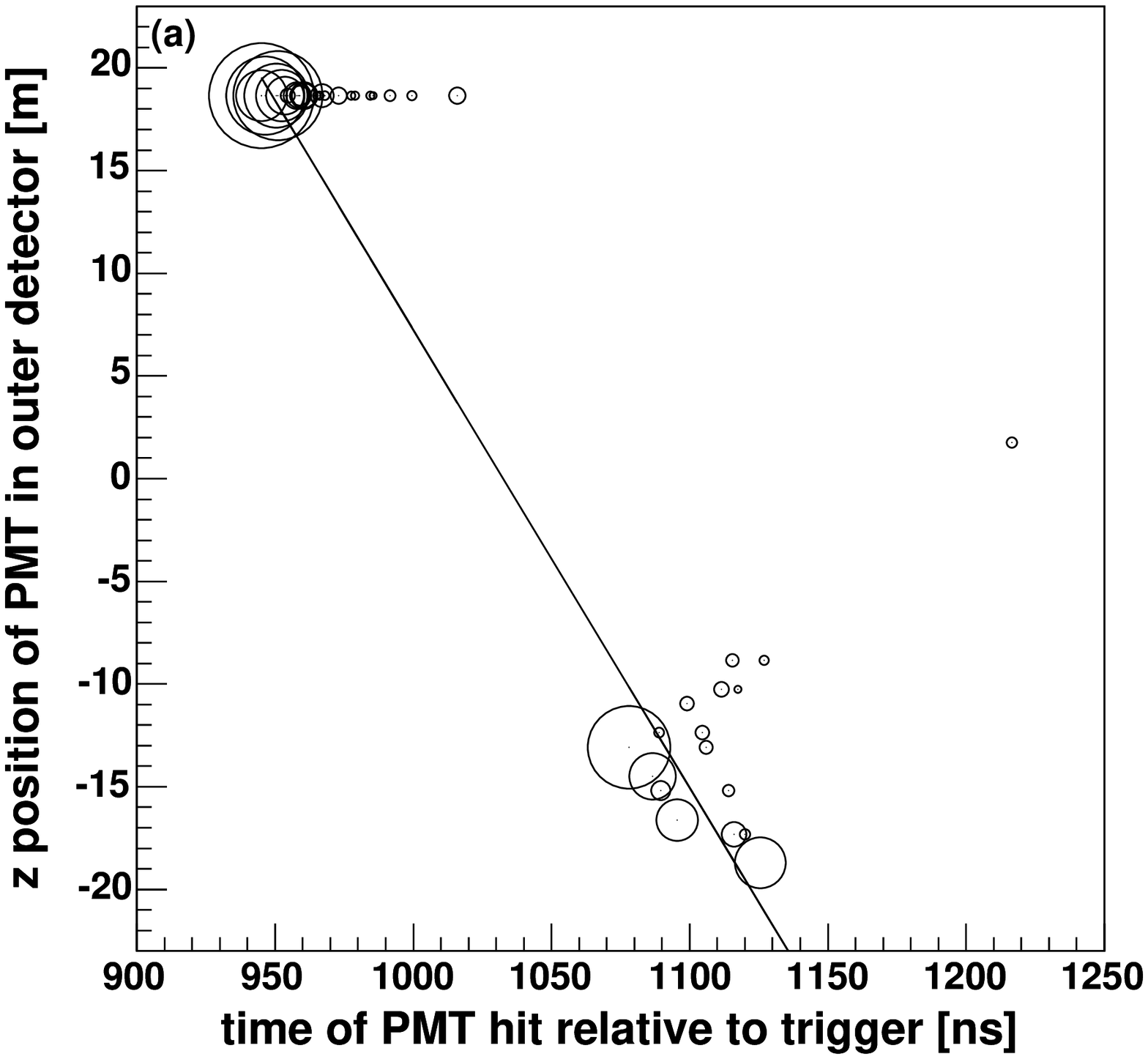}{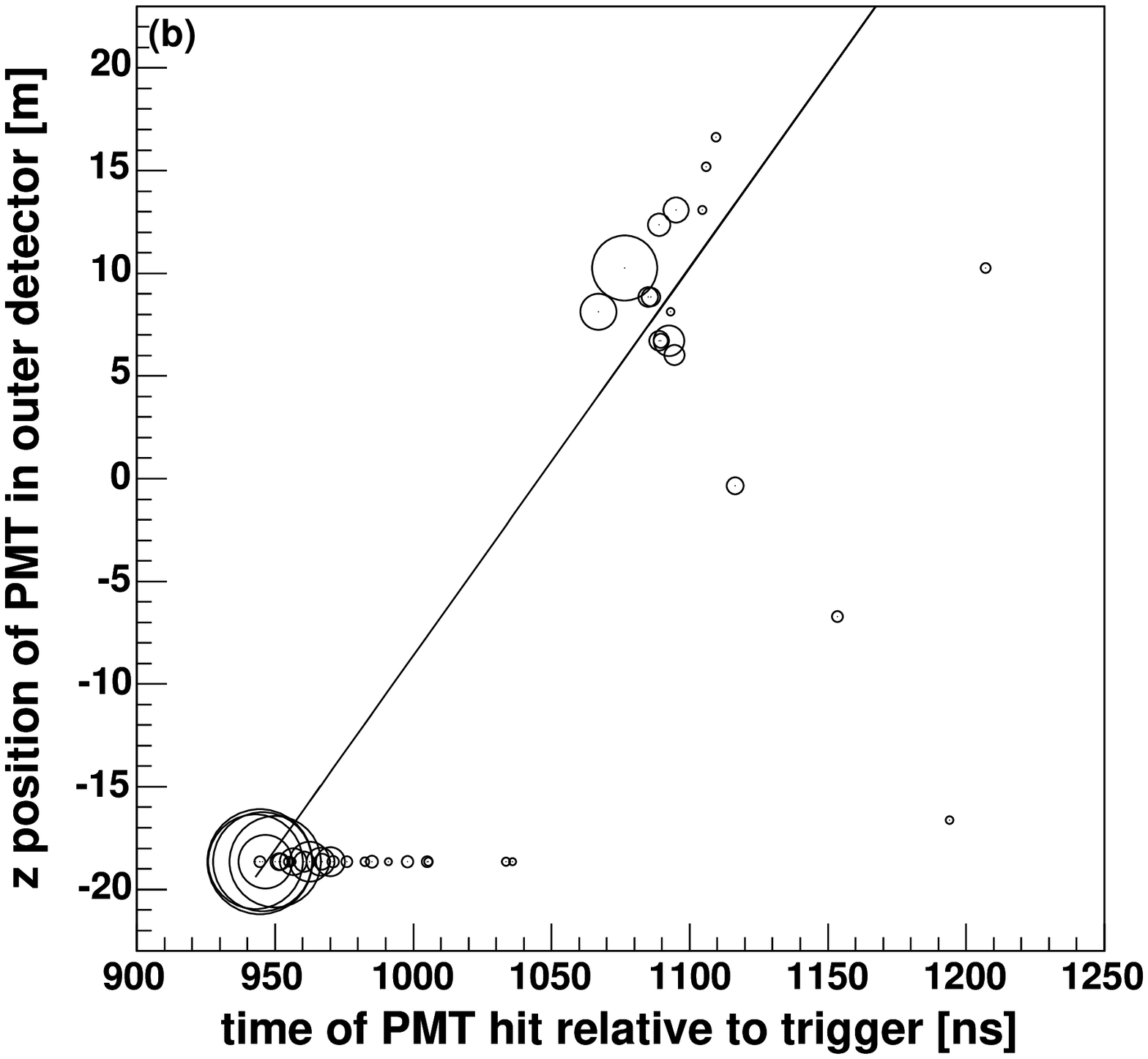}
\caption{
(\emph{a}) OD-based muon trajectory fit applied
to an example MC downward-going muon event. (\emph{b}) OD-based fit applied to an example
MC upward-going muon event. The size of the circle around each point is proportional
to the charge detected in the PMT.\label{fig:fit_examples}}
\end{figure*}

SK-I's ultra--high-energy data sample contains a total
of 52214 events. Most of these are either very energetic downward-going
cosmic-ray muons or multiple muon events where two or more downward-going
muons hit the detector simultaneously. In order to select candidate
upward-going muons from this sample, we applied a simple linear fit to the OD 
data for each event. A linear
fit was done on the $z$-position of each OD PMT versus the time it fired,
weighted by the total charge in the PMT. Example fits of simulated
downward-going and upward-going muon events
are shown in
Figure~\ref{fig:fit_examples}.%
The slope of this fitted line is an estimate of $-\cos{\Theta}$,
where $\Theta$ is the zenith angle of the incoming muon. A positive
fitted slope indicates that the muon is upward-going. A similar linear
fit was done on the $x$- and $y$-positions to determine the full muon trajectory
through the detector.

Since this fitting method is based on the OD (which has a lower
resolution than the ID), it is not as precise as the muon fitting algorithms
used in the lower energy upward-going muon analysis. However, it works even 
when the ID PMT electronics are completely saturated and the precision 
ID-based algorithms fail.

\subsection{Selection Cuts}

\label{sub:Selection-Cuts}

To select candidate upward-going muons, we applied 
the OD-based fit to all 52214
events in the ultra--high-energy data sample. A cut of $\cos{\Theta}\le0.1$
was used to eliminate the bulk of the downward-going single and multiple
muon events. In addition, the fitted trajectory was required to have a path
length of $>7{\rm {\; m}}$ in the ID.

To ensure high-quality fit results, we looked at the number of OD
PMTs hit near the projected entry and exit points. For a true
throughgoing muon, there will be a cluster of hit PMTs around the
entry and exit points. If the fit is accurate,
the projected path should pass through both of these clusters, so
we made an additional cut on the number of OD PMTs hit within $10{\rm {\; m}}$
of the projected OD entry and exit points, $N_{{\rm ODentry}}$
and $N_{{\rm ODexit}}$. We required both $N_{{\rm ODentry}}$
and $N_{{\rm ODexit}}$ to be 10 or greater.

Events that do not have $N_{{\rm ODentry}}$ and $N_{{\rm ODexit}}\ge10$ are generally either
stopping muons, partially contained events, or poorly fitted throughgoing muons.
Stopping muons are muons that stop in the detector and only form an OD entry cluster.  
They typically have energies of 1-10 GeV, well below the range of the 
expected astrophysical signal, so it
is appropriate to discard events that look like stopping muons.
Partially-contained events are neutrino interactions that take place inside the detector 
and only
form an OD exit cluster --- they are not part of the upward-going muon flux incident on the 
detector, so we want to discard these as well.
This cut also occasionally eliminates inaccurately fitted
throughgoing muon events, which
reduces the efficiency somewhat but improves the accuracy
of the fit results. 

Another possible type of event that can masquerade as a throughgoing
muon is a partially-contained event with multiple exiting particles.
Such an event will create two (or more) clusters of hit PMTs in the
OD, which could be mistaken as the entry and exit points of a throughgoing
muon. In order to eliminate such events, we looked at the timing between
the OD and ID entry points. If the event is truly a throughgoing muon,
the OD PMTs near the entry point should fire before the ID PMTs. We
determined OD and ID entry clusters via a simple time-based clustering
method and evaluated the mean time of hits within $6{\rm {\; m}}$
of the OD and ID entry points, $t_{{\rm IDentry}}$ and $t_{{\rm ODentry}}$.

In an ideal measurement, $t_{{\rm IDentry}}<t_{{\rm ODentry}}$
would indicate that the perceived entry cluster is actually caused
by an exiting particle. However, the timing determination is complicated
by an effect known as prepulsing.  
Prepulsing occurs when there are so many photons incident on the PMT that 
they are not all converted to photoelectrons at the photocathode - some photons
hit the first dynode instead and are converted to photoelectrons there.
When this happens
in an ID PMT, the PMT pulse
appears earlier relative to the OD light. 
If this occurs for 
an ultra-high energy throughgoing muon event, it could cause
 $t_{{\rm IDentry}}-t_{{\rm ODentry}}$ to be negative.
To allow for this effect, we used a fairly loose
cut of $-40{\rm {\; ns}}$: if $t_{{\rm IDentry}}-t_{{\rm ODentry}}<-40{\rm {\; ns}}$,
indicating very early ID light, then the event was rejected as a likely
neutrino interaction in the ID.

After these cuts on $\cos{\Theta}$, path length, $N_{{\rm ODentry}}$,
$N_{{\rm ODexit}}$, and $t_{{\rm IDentry}}-t_{{\rm ODentry}}$
were applied to the 52214 events in the sample, 343 candidate events
remained. These remaining events were then evaluated by a visual scan
and a manual direction fit by two independent researchers
to select events with $\cos{\Theta}<0$. 

The visual scan eliminates events that can pass the automatic reduction but are obvious to the trained human eye as noise, i.e., mainly downward-going multiple muon events and ``flashers.''  Multiple muon background events occur when two or more downward-going muons from an atmospheric shower pass through the detector simultaneously.  They have extra energy deposition (and thus are expected to be more common in the ultra--high-energy sample) and typically give poor OD fit results due to multiple OD clusters, but they are easy to identify visually.  Flashers are events caused by malfunctioning PMTs that emit light and create characteristic patterns.  After the multiple muon events and flashers are removed, the manual direction fit separates truly upward-going muons from mis-fitted downward-going muons.

From the 343 candidates, only one event passed the visual scan and manual direction
fit selection as being truly upward-going. The breakdown of the visual
scan and manual fit classifications is shown in Table~\ref{table:visual scan}.
\begin{table}
\begin{center}
\caption{Visual scan of candidate upward-going muon events
\label{table:visual scan}}
\begin{tabular}{cc}
\tableline 
\tableline
Visual scan classification&Number of events\\
\tableline 
Multiple muon events&164\\
PMT ``flashers'' (malfunctioning PMTs)&5\\
Other noise events&2\\
Downgoing muons (manual fit $\cos\Theta\ge0$)&171\\
Upgoing muons (manual fit $\cos\Theta<0$)&1\\
\tableline 
Total&343\\
\tableline
\end{tabular}
\end{center}
\end{table}

This upward-going muon event selected from the $\ge1.75\times10^{6}{\rm {\; pe}}$
sample is the ultra-high energy upward-going muon signal observed by SK-I.  This event
occurred on 2000 May 12 at 12:28:07 UT and deposited 1,804,716 pe in the ID.  
Based on the manual fit results, the path length through the ID was 40 m, and the 
zenith angle was $\cos{\Theta}=-0.63$, corresponding to a direction of origin of 
$({\rm{R.A.}}, {\rm{decl.}})=20^{\rm h}38^{\rm m},-37^{\circ}18\arcmin$.



\section{High-Energy Isotropic Monte Carlo}

\label{sec:High-Energy-Isotropic-MC}

In order to calculate the observed muon flux, we need to determine
the resolution and efficiency of the OD-based fit and other cuts on high-energy
muons, and we need to estimate the probability that a muon of
a given energy will deposit $\ge1.75\times10^{6}{\rm {\; pe}}$ in
the ID.

To determine these quantities, we generated a high-energy isotropic
Monte Carlo (MC) sample. This MC consists of an isotropic flux
of muons in monoenergetic bins impinging on the Super-K detector,
representing a flux of muons from neutrino interactions in the surrounding
rock. Seven monoenergetic bins were used, with muon energies ranging
from $100{\rm {\; GeV}}$ to $100{\rm {\; TeV}}$, and 10000 events
with a path length in the ID of $>7{\rm {\; m}}$ were generated in
each bin. The simulation was performed using a {\tt GEANT}-based detector
simulation. {\tt GEANT}'s muon propagation has been shown to agree with
theoretical predictions up to muon energies of $100{\rm {\; TeV}}$
\citep{2001NIMPA.459..319B,2003ICRC....3.1673D}.

\subsection{Resolution and Efficiency of Event Selection}

\label{sub:Resolution-and-Efficiency}

\begin{figure}
\plotone{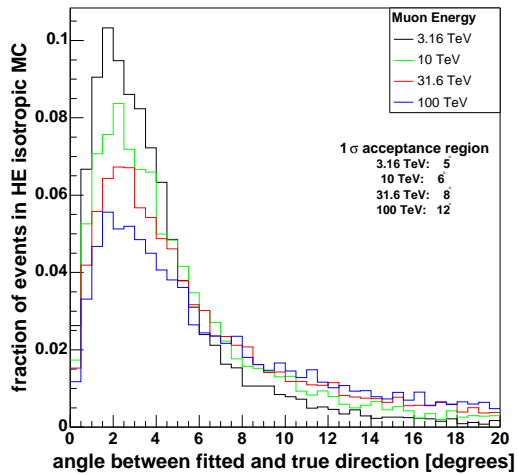}
\caption{Angular resolution of OD-based fit for events 
from the high energy isotropic MC with $>10^{5}{\rm {\; pe}}$
in the ID.\label{fig:angle_rock}}
\end{figure}

The muon trajectory fitting algorithm
 discussed in \S~\ref{sub:Linear-Fit-Method}
was applied to the high-energy isotropic MC. A plot 
illustrating the angular resolution
of the fit is shown in Figure~\ref{fig:angle_rock}.
The resolution of 5--12$^\circ$ is poor compared to the typical $1^\circ$ 
resolution of precision ID fitting algorithms, but those
algorithms do not work well on
these high-energy saturated events. 

The efficiency of the upward-going cuts was estimated by considering all of 
the MC events with true
values of $\cos{\Theta}<0$ and ID path length $>7{\rm {\; m}}$
and then determining the fraction of these events that pass the selection 
cuts described in \S~\ref{sub:Selection-Cuts}. This was done using 
isotropic MC events
with $>10^{5}{\rm {\; pe}}$ since there were relatively few events in the MC with 
$\ge1.75\times10^{6}{\rm {\; pe}}$. (Above $10^{5}{\rm {\; pe}}$ the efficiency and 
resolution do not depend strongly on the number of ID pe deposited.) Also, as 
discussed in \S~\ref{sub:High-Energy-Fraction},
only the energy bins in the range $3.16-100{\rm {\; TeV}}$ were considered.
The efficiency
was calculated as a function of muon energy and $\cos\Theta$. 

Statistical uncertainties on the efficiency determination were calculated
using the Bayesian method discussed by 
\citet{Conway:2002}.
Systematic
uncertainties due to uncertainties in the fitted values of the zenith
angle, path length, and number of OD PMTs hit near the entry and exit
points were evaluated by comparing the fitted values with the true MC values.  The fitted
values agree well with the MC values, and distributions of the difference
between the fitted values and the MC values were used to determine an effective
$1\,\sigma$ region for each parameter. The cuts on these parameters were varied 
by $1\,\sigma$ in either direction to determine the effect on the efficiency.


Another source of systematic uncertainty on the efficiency is 
prepulsing, which is not included in the MC simulation and must be estimated
separately.
To do this, we compared the results from
the high-energy isotropic MC to a sample of 627 ultra--high-energy 
downward-going muon data events with $\ge1.75\times10^{6}{\rm {\; pe}}$ in the ID
selected by a visual scan. For many of these data events,
the time difference $t_{{\rm IDentry}}-t_{{\rm ODentry}}$ is negative, showing 
evidence of prepulsing not seen in the MC sample. 

To estimate the systematic
uncertainty on the efficiency calculation due to prepulsing, we calculated 
a heuristic correction to the MC by smearing the MC distribution of $t_{{\rm 
IDentry}}-t_{{\rm ODentry}}$ such that it matched the distribution of the 
ultra--high-energy downward-going data.
We applied this adjustment to the events in the MC sample and found that an 
additional 1\% of the MC events would be cut after accounting for prepulsing, which 
lengthens the negative-side error bars on the efficiency estimates by approximately 0.01. 

Finally, one additional correction must be made: the efficiency has
been estimated \emph{at} different values for $E_{\mu}$, but the
flux calculation is for the flux \emph{above} a threshold energy $E_{\mu}^{\rm min}$.
Since the efficiency decreases with energy, $\varepsilon\left(E_{\mu},\Theta\right)$
is an overestimate for $\varepsilon\left(\ge E_{\mu}^{\rm min},\Theta\right)$.
This does \emph{not} lead to a conservative upper limit for the flux, so
a correction must be made.  This requires knowledge of the energy spectrum
of the expected signal, so we modeled the signal as an isotropic flux
of neutrinos with ${d\Phi_{\nu}\left(E_{\nu}\right)}/{dE_{\nu}}\propto E_{\nu}^{-2}$
(a plausible astrophysical spectrum \citealt{1990cup..book.....G}), used the method of
 \S~\ref{sec:Analytical-Estimate} to estimate the muon flux
$\Phi_{\mu}\left(\geq E_{\mu}^{\rm min}\right)$, and used this muon flux 
to extrapolate our efficiency calculations to find 
$\varepsilon\left(\ge E_{\mu}^{\rm min},\Theta\right)$.
%
%
%
%
%
This procedure yields small downward adjustments ($0.6-9\%$) to the calculated
efficiency in each angular bin. 

Additionally, we must also account for the efficiency of the visual
scan and manual fit procedure. To do this, we did a visual scan of
the 605 events from the high-energy isotropic MC with
true $\cos{\Theta}<0$, true ID path length $>7{\rm {\; m}}$, and
$\ge1.75\times10^{6}{\rm {\; pe}}$ in the ID and found that 10 of
these events were eliminated in the visual scan. This gives an efficiency
of roughly $98\%$, so we adjust our efficiency results by a factor
of $0.98$. 

\begin{figure}
\plotone{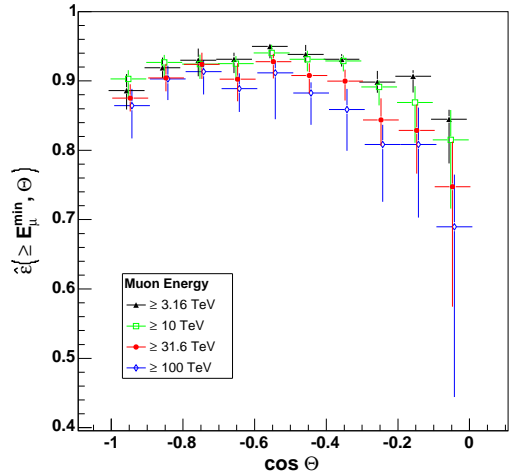}
\caption{Efficiency of our data reduction on 
upward, throughgoing, $\ge 1.75\times10^{6}{\rm {\; pe}}$ muons  
as determined by the high-energy isotropic MC. Error
bars include statistical and systematic errors.\label{fig:eff_total}}
\end{figure}

The final results for the efficiency of our cuts on upward, throughgoing, $\ge1.75\times10^{6}{\rm {\; pe}}$ muons (including all corrections
discussed above) are plotted in Figure~\ref{fig:eff_total}.%

\subsection{Ultra--High-Energy Fraction}

\label{sub:High-Energy-Fraction}

In order to set an upper limit on the flux of upward-going
muons from cosmic neutrinos one must make an inference about the energies
of the upward-going muons in the ultra--high-energy sample. Above energies
of $\sim1{\rm {\; TeV}}$, muon energy loss in water is dominated
by radiative processes such as bremsstrahlung,  
so high-energy muons
have some probability of depositing large numbers of photoelectrons
in the Super-K detector and contributing to the ultra--high-energy sample.
Since this energy loss is not continuous, it is not possible to estimate
the muon energy for a single ultra--high-energy upward-going muon event.
Rather, MC is used to make a statistical statement about
the energies of the muons that make up the $\ge1.75\times10^{6}{\rm {\; pe}}$
sample.

The high-energy isotropic MC has been used to determine the
fraction $k\left(E_{\mu}\right)$ of muons with energy $E_{\mu}$
that will deposit $\ge1.75\times10^{6}{\rm {\; pe}}$ in the ID, thus
contributing to the ultra--high-energy sample. Results are shown in
Table~\ref{table:highe_frac}.%
\begin{table}
\begin{center}
\caption{Fraction of high-pe events $k\left(E_{\mu}\right)$
in high energy isotropic MC\label{table:highe_frac}}
\begin{tabular}{ccc}
\tableline 
\tableline
&
number of MC events&
$k\left(E_{\mu}\right)$ \\
$E_{\mu}$&
(out of 10000) with&
(with statistical\\
(TeV)&
$\ge1.75\times10^{6}{\rm {\; pe}}$ in ID&
uncertainties)\\
\tableline
0.1, 0.316, 1&
0&
$0.0000\begin{array}{c}
+0.0001\\
-0.0000\end{array}$\\
$3.16$&
35&
$0.0035\pm0.0006$\\
$10$&
113&
$0.0113\begin{array}{c}
+0.0011\\
-0.0010\end{array}$\\
$31.6$&
293&
$0.0293\pm0.0017$\\
$100$&
879&
$0.0879\begin{array}{c}
+0.0029\\
-0.0028\end{array}$\\
\tableline
\end{tabular}
\end{center}
\end{table}
 Statistical uncertainties were calculated using the Bayesian method
discussed by \citet{Conway:2002}. As can be seen in Table~\ref{table:highe_frac},
the three lowest energy bins --- from $100\;{\rm {GeV}}$ to $1\;{\rm {TeV}}$
--- do not make a significant contribution to the $\ge1.75\times10^{6}{\rm {\; pe}}$
sample. Hence, the rest of this analysis was done using only the four
highest energy bins --- from $3.16$ to $100\;{\rm {TeV}}$.

These results for $k\left(E_{\mu}\right)$ are used in \S~\ref{sec:Flux-Calculation}
to calculate the flux $\Phi_{\mu}\left(\geq E_{\mu}^{\rm min}\right)$.
Since $k\left(E_{\mu}\right)$ increases with energy, it is an underestimate
for $k\left(\geq E_{\mu}^{\rm min}\right)$, which leads to a conservative
upper limit for $\Phi_{\mu}\left(\geq E_{\mu}^{\rm min}\right)$. 

Also, since this MC only simulates the muon and not the actual
neutrino interaction, the effect of lower energy debris from deep inelastic
scattering events that make it into the detector has been neglected,
again leading to an underestimate of $k\left(E_{\mu}\right)$. This
effect is expected to be small in the energy range considered here
--- above $E_{\mu}=1\;{\rm {TeV}}$ at the detector entry, 
over 80\% of the neutrino-induced upward-going
muons come from over $200\;{\rm{m}}$ away from the detector, so most of
the debris is absorbed by the surrounding rock, as
determined using the atmospheric neutrino MC discussed in
\S~\ref{sec:Expected-Atmospheric-Background}.


\section{Flux Calculation}

\label{sec:Flux-Calculation}

The flux of upward-going muons above a threshold energy $E_{\mu}^{\rm min}$
is given by
\begin{equation}
\Phi_{\mu}\left(\geq E_{\mu}^{\rm min}\right)=\frac{1}{2\pi T k\left(\geq E_{\mu}^{\rm min}\right)}{\displaystyle \sum\limits _{j=1}^{n}\frac{1}{\varepsilon\left(\ge E_{\mu}^{\rm min},\Theta_{j}\right) A\left(\Theta_{j}\right)}},\label{eq:flux}\end{equation}
where $n$ is the total number of upward-going muon events observed and
$\Theta_{j}$ is the zenith angle of the $j$th event. The efficiency
$\varepsilon\left(\ge E_{\mu}^{\rm min},\Theta_{j}\right)$ and the ultra--high-energy
fraction $k\left(\geq E_{\mu}^{\rm min}\right)$ are calculated in \S~\ref{sec:High-Energy-Isotropic-MC}.
$T$ is the detector livetime, which is 1679.6 days for SK-I. $A\left(\Theta_{j}\right)$
is the effective area of the Super-K detector perpendicular to the direction
of incidence for tracks with a path length of $>7{\rm {\; m}}$ in
the ID. The average effective area of the detector is $\sim 1200{\rm {\; m}}^{2}$.

Equation~(\ref{eq:flux}) has been applied to the detected upward-going muon
event discussed in \S~\ref{sub:Selection-Cuts} to calculate $\Phi_{\mu}\left(\geq E_{\mu}^{\rm min}\right)$
for $E_{\mu}^{\rm min}$ in the range $3.16-100{\rm {\; TeV}}$. Results
are shown in Table~\ref{table:flux_results}.%
\begin{table}
\begin{center}
\caption{The flux of ultra-high energy upward-going
muons as observed by SK-I.\label{table:flux_results} }
\begin{tabular}{cc}
\tableline 
\tableline
$E_{\mu}^{\rm min}$&
$\Phi_{\mu}\left(\geq E_{\mu}^{\rm min}\right)$\\
(TeV)&
(${\rm {cm^{-2}s^{-1}sr^{-1}}}$)\\
\tableline
$3.16$&
$2.64\times10^{-14}\begin{array}{c}
+16.1\%\\
-17.9\%\end{array}$\\
$10$&
$8.23\times10^{-15}\begin{array}{c}
+9.48\%\\
-9.73\%\end{array}$\\
$31.6$&
$3.25\times10^{-15}\begin{array}{c}
+6.67\%\\
-6.40\%\end{array}$\\
$100$&
$1.10\times10^{-15}\begin{array}{c}
+4.96\%\\
-4.09\%\end{array}$\\
\tableline
\end{tabular}
\tablecomments{These fluxes include both atmospheric background and potential 
astrophysical signal at each threshold energy.}
\end{center}
\end{table}
 Systematic uncertainties include a $0.1\%$ uncertainty on the live time
$T$, a $0.3\%$ uncertainty on the effective area $A$, the total
efficiency uncertainties shown in Figure~\ref{fig:eff_total}, and
the statistical uncertainties on $k$ shown in Table~\ref{table:highe_frac}.
This flux includes both the potential signal from astrophysical neutrinos
and a background of atmospheric neutrinos.

\section{Expected Atmospheric Background from Monte Carlo}

\label{sec:Expected-Atmospheric-Background}

When searching for neutrinos from astrophysical sources, the dominant
background is the atmospheric neutrino spectrum.
Atmospheric neutrinos are produced by decays of pions and kaons formed
when cosmic rays interact with particles in the atmosphere. 
We have used an atmospheric neutrino MC that
is a 100 yr equivalent sample of events
due to the atmospheric neutrino flux.
The neutrino flux in \citet{2004PhRvD..70d3008H} was used up to neutrino
energies of 1 TeV. At 1 TeV, the calculated flux in \citet{1980SvJNP..31..784V}
was rescaled to the Honda et al. flux. Above 1 TeV, the
rescaled flux from Volkova was used up to 100 TeV. Neutrino
interactions were modeled using the GRV94 parton distribution functions
\citep{1995ZPhyC..67..433G}, and muon propagation through the rock and water was modeled 
using {\tt GEANT}. Further details on the atmospheric neutrino MC can
be found in \citet{2005PhRvD..71k2005A}.
No correction is made for neutrino oscillations, because based on the 
oscillation parameters determined in \citet{2005PhRvD..71k2005A}, the neutrino 
oscillation probability is negligible for neutrinos above $1{\rm {\; TeV}}$.

This atmospheric MC is
split into two parts: a partially-contained/fully-contained (PC/FC)
sample, which consists of events with neutrino interaction points
inside the ID plus a shell 50 cm thick surrounding the ID (the insensitive
region), and an upward-going muon sample, which consists of
events with neutrino interaction points outside the ID. Note that
these two samples overlap because they both cover the 50 cm insensitive
region.

The OD-based fit was applied to the events in the atmospheric MC, using
the same cuts that were applied to the SK-I data. A total of 11 MC events passed the 
$\ge1.75\times10^{6}{\rm {\; pe}}$, $\cos{\Theta}\le0.1$,
path length $>7{\rm {\; m}}$, $N_{{\rm ODentry}}$ and $N_{{\rm ODexit}}\ge10$,
OD/ID timing, and manual fit cuts. Out of these 11, 2 are from the
PC/FC sample, both with interaction points inside the ID. The remaining
9 events are from the upward-going muon sample: 3 events with interaction points
in the 50 cm insensitive region, 1 event in the water of the OD, and
5 events in the rock surrounding the detector. 

All of these background events are deep inelastic scattering (DIS) events where an interaction between a muon neutrino and a nucleon produces a muon plus a spray of lower energy particles.  The 6 events with interaction points within the detector (ID or OD) have muon energies of $0.1-0.8{\rm {\; TeV}}$, and the 5 events occurring in the rock have muon energies of $1-20{\rm {\; TeV}}$.  This difference in the energy range can be understood as follows: For DIS events occurring a long distance ($>2{\rm {\; m}}$ or so) from the detector, only the muon will reach the detector since the lower energy debris will be absorbed by the surrounding rock, but for nearby events or events occurring in the water of the OD, some of these lower energy particles will enter the detector as well.  This means that nearby events can be included in the $\ge1.75\times10^{6}{\rm {\; pe}}$ sample with lower muon energies than more distant events.    

Since the insensitive region is covered by both the PC/FC and
the upward-going muon MC samples, we divided the 3 events originating from
this region in half, for a total of 1.5 events in the insensitive
region. This gives a total of 9.5 MC events in 100 yr
of simulated live time. Scaling the 100 yr MC to SK-I's
live time of 1679.6 days gives an expected background of $0.44$ events
due to atmospheric neutrinos during the operation of SK-I.

The statistical uncertainty in this background measurement
of the MC events is 31\%.
There are also significant systematic uncertainties:
the normalization of the atmospheric neutrino flux has a theoretical
uncertainty of $\pm10\%$ at neutrino energies below $10{\rm {\; GeV}}$
\citep{2005PhRvD..71k2005A}. In order to extend this to the energy range of the
expected background, we must also account for the uncertainty of 0.05 in the spectral index 
of the primary cosmic ray spectrum above 100 GeV, which leads to a 0.05 uncertainty
in the spectral index for atmospheric neutrinos above $10{\rm {\; GeV}}$
\citep{2005PhRvD..71k2005A}. 

To determine how much this uncertainty affects our
result for the background, we consider the atmospheric neutrino flux
to be known at $10{\rm {\; GeV}}$, and we calculate
the uncertainty of the total flux $\Phi_{\nu}$ above a threshold
energy $10.6{\rm {\; TeV}}$, the average neutrino energy
of the atmospheric MC events passing our cuts.  For a differential
flux of ${d\Phi_{\nu}}/{dE_{\nu}}\propto E_{\nu}^{-\gamma}$ with $\gamma=3.7\pm0.05$,
the spectral index uncertainty gives us a
$\pm37\%$ uncertainty on the atmospheric neutrino flux $\Phi_{\nu}$.


Finally, the
neutrino cross-section at high energies is thought to be known to
within $10\%$ or less, so we include an additional $10\%$ uncertainty
to account for this. These uncertainties are summarized in Table~\ref{Table:bg_sys}
and lead to a total uncertainty on the background of $50\%$.

\begin{table}
\begin{center}
\caption{Systematic uncertainties in atmospheric
neutrino background\label{Table:bg_sys}}
\begin{tabular}{cc}
\tableline 
\tableline
Source of Uncertainty&
Uncertainty\\
\tableline
Statistical&
31\%\\
Absolute normalization of atm $\nu$ flux&
10\%\\
Primary spectral index&
37\%\\
Neutrino cross-section uncertainty&
10\%\\
\tableline 
Total uncertainty in background flux&
50\%\\
\tableline
\end{tabular}
\end{center}
\end{table}

 Other potential errors (uncertainties
in the simulation of the SK detector and varying hadron multiplicities
in different deep inelastic scattering models) were tested and shown
to not make a significant contribution to the systematic uncertainty. 

Another potential background source of high-energy neutrinos not included
in the 100 yr atmospheric MC is the prompt atmospheric
neutrino flux, which arises from decays of short-lived charmed particles
produced when cosmic rays interact with particles in the atmosphere.
This flux is not as well-understood as the conventional atmospheric
neutrino flux from decays of pions and kaons, but it is expected to
have a harder spectrum and therefore is expected to become more important
as we push towards higher energy scales. 

Based on the 100 yr atmospheric MC, we calculate
an expected background for this analysis of $0.44\pm0.22$ events,
compared to the 1 event observed. However, there are three effects
that this MC does not take into account: it does not
include neutrinos over $100{\rm {\; TeV}}$, it does not include the
prompt atmospheric neutrino flux, and it does not account for attenuation
of neutrinos passing through the Earth. We account for these
issues by making corrections based on an analytical calculation discussed
in \S~\ref{sec:Analytical-Estimate}.

Also, it is important to note 
that the atmospheric background --- both conventional and prompt
--- comes from a lower energy range than that for which we expect to observe
the possible signal of neutrinos from astrophysical sources. Roughly speaking,
the peak 90\% of the expected upward-going muon events 
come from muons with energies in the range $E_{\mu}=0.02-10{\rm \; TeV}$
for conventional atmospheric neutrinos, and $E_{\mu}=0.2-200{\rm \; TeV}$
for prompt neutrinos. As we learned from the 11 background events
in the 100 yr atmospheric MC, 
muons with energies
 $E_{\mu}<1{\rm \; TeV}$ contribute to the  $1.75\times10^{6}{\rm {\; pe}}$ sample mainly via debris from DIS events very close to the detector rather than from catastrophic energy loss of the muon. 

Since there are many more low-energy events in the atmospheric spectrum, 
they will dominate
even though each one only has a tiny probability of depositing a large
amount of energy in the detector. In contrast, the peak 90\% of the
expected muon events from a harder spectrum --- a hypothetical $E_{\nu}^{-2}$
astrophysical flux --- come from the range $E_{\mu}=7-6000{\rm \; TeV}$.
Thus, even though the atmospheric flux is very small in the energy
range $E_{\mu}=3-100{\rm \; TeV}$ and above where we are setting
our limit, the high-pe tails of the distribution from lower energy
events dominate our background simply because of the much larger flux
of lower energy atmospheric events.

\section{Analytical Estimate of Expected Muon Flux}

\label{sec:Analytical-Estimate}

\subsection{Method for Calculating Muon Flux}

\label{sub:Method-for-Calculating-Muon-Flux}

In order to better understand our observed flux of high-energy upward-going
muons, we developed a method to calculate the expected upward-going muon
event rate due to a predicted flux of neutrinos. We have used this
to plot curves for theoretical muon fluxes in Figure~\ref{fig:limits} and also to adjust
the atmospheric background calculated with MC in \S~\ref{sec:Expected-Atmospheric-Background}
by correcting for effects not included in the simulation. 

To convert a model neutrino flux into an expected upward-going muon event
rate, we follow the calculation detailed in \citet{1995PhR...258..173G} and
 \citet{1996APh.....5...81G}.
The flux of muons $\Phi_{\mu}\left(\geq E_{\mu}^{\rm min}\right)$ above
an energy threshold $E_{\mu}^{\rm min}$ is given by

\begin{equation}
\Phi_{\mu}\left(\geq E_{\mu}^{\rm min}\right)=\int_{E_{\mu}^{\rm min}}^{\infty}dE_{\nu}P_{\mu}\left(E_{\nu},\, E_{\mu}^{\rm min}\right)\frac{d\Phi_{\nu}^{\rm av}\left(E_{\nu}\right)}{dE_{\nu}},\label{eq:muflux}\end{equation}
where $P_{\mu}\left(E_{\nu},\, E_{\mu}^{\rm min}\right)$ is the probability
that an incoming neutrino with energy $E_{\nu}$ will produce a muon
with energy above the threshold $E_{\mu}^{\rm min}$ at the detector,
and ${d\Phi_{\nu}^{\rm av}\left(E_{\nu}\right)}/{dE_{\nu}}$ is the
differential neutrino flux averaged over solid angle and reduced by an exponential
factor due to attenuation of the neutrinos as they pass through the Earth. 

$P_{\mu}\left(E_{\nu},\, E_{\mu}^{\rm min}\right)$
depends on both the charged-current neutrino cross-section and the
energy lost by the resulting muon as it propagates through the
Earth. 
%
%
%
In our calculation,
we calculate the cross-section using the GRV94 parton distribution
functions \citep{1995ZPhyC..67..433G} applied with the code used in \citet{1996APh.....5...81G} provided 
by M.~Reno (2005, private communication),
and we use an effective muon range calculated
using MC methods \citep{1991PhRvD..44.3543L}.

As discussed in \citet{1996APh.....5...81G}, the appropriate cross-section 
$\sigma\left(E_{\nu}\right)$ to use in the neutrino attenuation 
factor lies between the charged current cross section and the sum of the charged and 
neutral current cross sections, so we have calculated the flux with the upper and
lower limits and averaged the resulting fluxes together.



We integrated equation~(\ref{eq:muflux}) 
numerically using various predicted models of the neutrino
flux ${d\Phi_{\nu}}/{dE_{\nu}}$ to calculate theoretical 
$\Phi_{\mu}\left(\geq E_{\mu}^{\rm min}\right)$ curves.  These are shown in 
Figure~\ref{fig:limits} for comparison to our experimental limits. 

%

\subsection{Analytical Estimate of Expected Background}

\label{sub:Analytical-Background}

Of particular interest here are the model fluxes of the background
due to atmospheric neutrinos, both conventional and prompt. We
use the method discussed in \S~\ref{sub:Method-for-Calculating-Muon-Flux} to correct 
for the omissions in the 100 yr atmospheric MC
discussed in \S~\ref{sec:Expected-Atmospheric-Background} by accounting for
neutrinos over $100{\rm {\; TeV}}$, attenuation of neutrinos in the
Earth, and the flux of prompt neutrinos.

\begin{figure}
\plotone{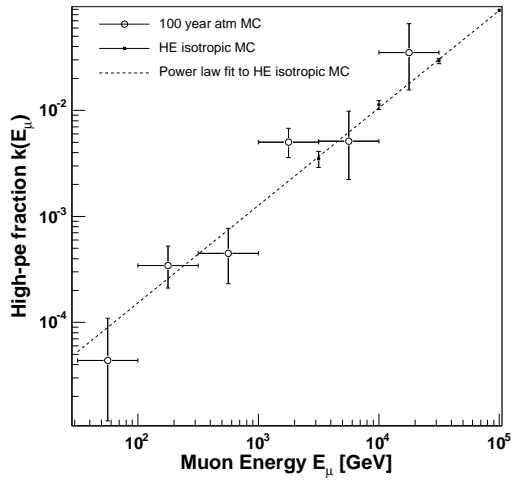}
\caption{Fraction of events $k\left(E_{\mu}\right)$
that deposit $\ge1.75\times10^{6}{\rm {\; pe}}$ in the ID
for the 100 yr atmospheric MC (\emph{open circles}), the high-energy isotropic
MC (\emph{filled squares}), and a power-law fit to the results from high-energy isotropic
MC. The power law is a reasonable approximation for the dominant
energy range of the atmospheric flux ($E_{\mu}=100{\rm \; GeV}-30{\rm \; TeV}$).\label{fig:kfit}}
\end{figure}

The expected number of events $N$ seen by Super-K in livetime $T$
is given by

\begin{equation}
N=2\pi TA_{\rm av}\int_{0}^{\infty}dE_{\mu}^{\rm min}\frac{d\Phi_{\mu}\left(\geq E_{\mu}^{\rm min}\right)}{dE_{\mu}^{\rm min}}k\left(E_{\mu}^{\rm min}\right),\label{eq:sk_background}\end{equation}
where ${d\Phi_{\mu}\left(\geq E_{\mu}^{\rm min}\right)}/{dE_{\mu}^{\rm min}}$
is the derivative of the curve calculated by the method in \S~\ref{sub:Method-for-Calculating-Muon-Flux}
and $k\left(E_{\mu}^{\rm min}\right)$ is the fraction of muons with energy
above $E_{\mu}^{\rm min}$ that will deposit $\ge1.75\times10^{6}{\rm {\; pe}}$
in the ID. 
We estimated $k\left(E_{\mu}^{\rm min}\right)$ using the results of the
high-energy isotropic MC discussed in \S~\ref{sub:High-Energy-Fraction}
by fitting the data in Table~\ref{table:highe_frac} with
a simple power law with a cutoff at 1 since $k$ represents a probability.
%
%
Figure~\ref{fig:kfit} compares this power-law fit to results 
from 100 yr atmospheric MC, illustrating that the power law is reasonable
in the energy range we are considering.%

We analytically estimated the expected
number of events in the 100 yr atmospheric MC by
starting with the same input neutrino flux,
applying equation~(\ref{eq:muflux}) \emph{without}
the exponential neutrino attenuation factor and integrating up to a maximum 
neutrino energy of $E_{\nu}^{\rm max}=100{\rm \; TeV}$.
We then used this muon flux in equation~(\ref{eq:sk_background}) to find
the expected number of background events in 100 yr and obtained
 $N_{100}=7.39$ expected events. (The 100 subscript denotes 100 yr of exposure.)
This matches our results from the 100 yr atmospheric MC to within 
statistical uncertainties.

To estimate the effects
of neutrinos above 100 TeV, we repeated the above calcuation 
without the $E_{\nu}^{\rm max}=100{\rm \; TeV}$ cutoff 
and obtained $N_{100}=7.47{\rm \; events}$, 
an increase of $1.1\%$.
Including the neutrino attenuation factor as well, we obtained
$N_{100}=7.29\pm0.02{\rm \; events}$ per 100 yr of exposure, 
a decrease of $2.4\%\pm0.3\%$. 
(The uncertainty is due to the choice of cross-section.) 


We also used this calculation to correct for the prompt atmospheric neutrino 
flux.  To account for the theoretical uncertainty in the prompt flux due
to differences between various flux models, we defined a 
high model and a low model for the prompt flux that bracket the models shown 
in Figure~1 of \citet{2003PhRvD..67a7301G} that are not ruled out by 
experimental limits.
The models used are discussed in more
detail in \citet{1996APh.....5..309T}, \citet{1993APh.....1..297Z}, \citet{2002NuPhS.110..531R}, \citet{1989NCimC..12...41B}, \citet{1999PhRvD..59c4020P}, and \citet{2000PhRvD..61e6011G, 2000PhRvD..61c6005G}.
Our analytical calculation 
gives $N_{100}=0.033{\rm \; events}$  
for the low model and $N_{100}=0.94{\rm \; events}$
for the high model, corresponding to a prompt flux that is $6.7\%\pm6.2\%$ 
of the flux of conventional atmospheric neutrinos.

Based on these results, we correct the background estimate made using 
the 100 yr atmospheric MC in
\S~\ref{sec:Expected-Atmospheric-Background} by 
applying these relative scalings to the result of $0.44{\rm \; events}$
in the SK-I livetime from \S~\ref{sec:Expected-Atmospheric-Background}.
This gives us a final result for the background of 
$0.46\pm0.23{\rm \; events}$ for the SK-I exposure. 


\section{Upper Limit for Muon Flux from Cosmic Neutrinos}

\label{sec:Upper-Limit}

Using the observed ultra--high-energy upward-going muon signal of 1 event
and the expected atmospheric neutrino background of $0.46\pm0.23$
events, we have calculated $90\%$ confidence upper limits for the
upward-going muon flux in the $3.16-100{\rm {\; TeV}}$ range due to neutrinos
from astrophysical sources (or any other non-atmospheric sources). 

This was done using the method of \citet{1998PhRvD..57.3873F}, with the
systematic uncertainties incorporated using the method of \citet{1992NIMPA.320..331C}, as implemented by \citet{2003PhRvD..67a2002C}
and improved by \citet{2003PhRvD..67k8101H}. This method incorporates both uncertainties
in the background flux and uncertainties in the flux factor $f$ relating
the observed number of events $n$ to the observed flux $\Phi$: $\Phi=f\, n$.
The uncertainty in $f$ includes systematic errors in the livetime,
effective area, efficiency, and ultra--high-energy fraction. For the confidence
interval calculation, the largest percent error for each energy bin
from Table~\ref{table:flux_results} was used as the percent uncertainty
in $f$. The uncertainties in both the background and the flux factor
were assumed to have a Gaussian distribution.

\begin{figure}
\plotone{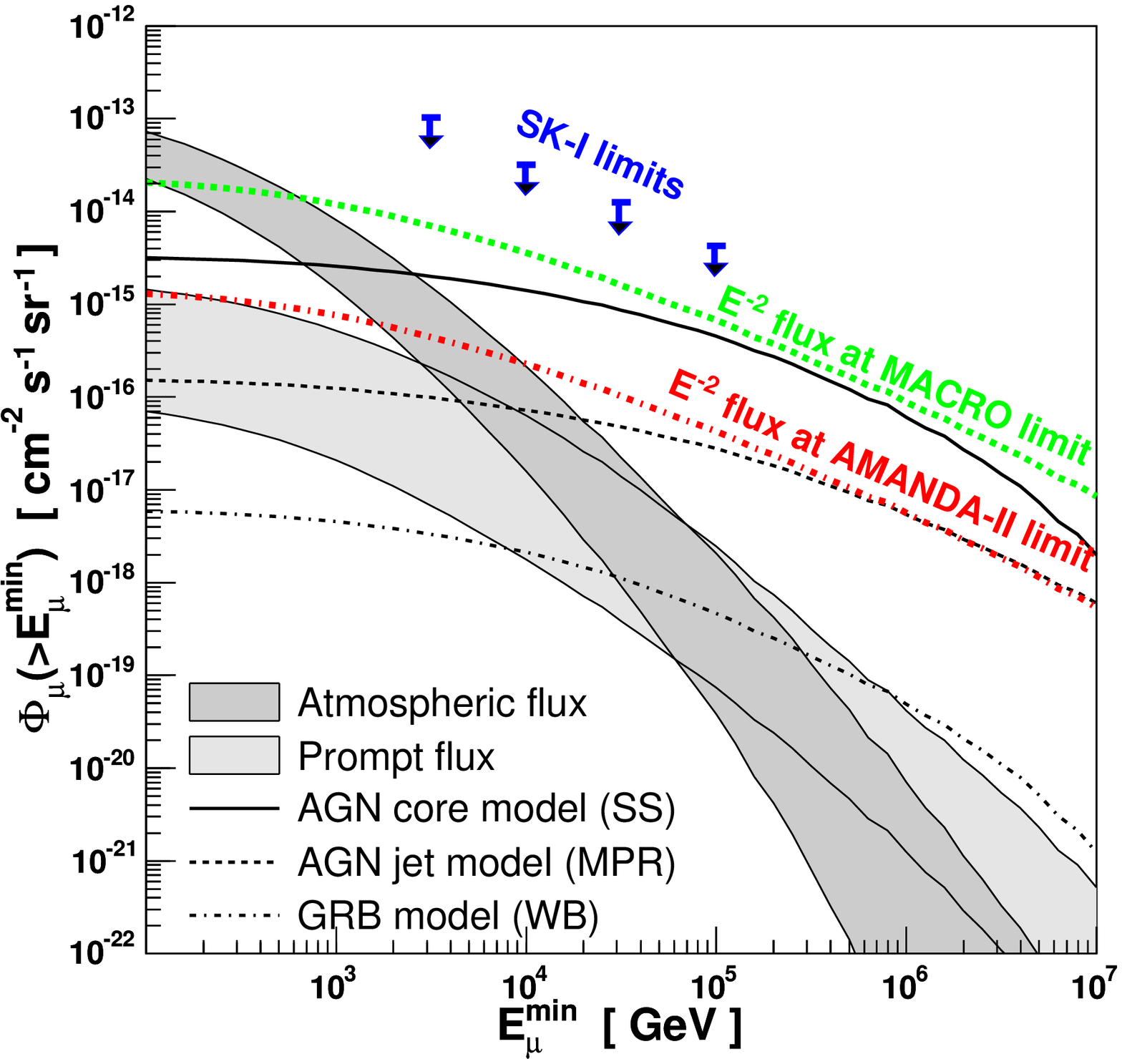}
\caption{Upper limits from this analysis on muon $\left(\mu^{+}+\mu^{-}\right)$
flux above energy threshold $E_{\mu}^{\rm min}$, compared to various
model fluxes. Models shown for muon flux due to astrophysical neutrinos
are AGN models from SS \citep{Stecker:1995th} and
MPR \citep{Mannheim:1998wp}, and a GRB model from WB \citep{1999PhRvD..59b3002W}.
Also shown is the atmospheric background, as modeled by
\citet{2004PhRvD..70d3008H} below $E_{\nu}=1{\;{\rm TeV}}$ and by \citet{1980SvJNP..31..784V}
rescaled to match the Honda model above $E_{\nu}=1{\;{\rm TeV}}$.
The upper edge of the atmospheric band represents the horizontal flux,
and the lower edge represents the vertical flux. The background due
to muons from prompt atmospheric neutrinos (assumed to be isotropic)
is shown for a range of possible models
as summarized in \citet{2003PhRvD..67a7301G}. Finally, we also show the expected
muon flux from a model neutrino flux that is isotropic and proportional
to $E_{\nu}^{-2}$ at the values of the limit set by MACRO \citep{2003APh....19....1M}
and AMANDA-II \citep{2005astro.ph..5278G}. The models of the neutrino flux have
been converted to a muon flux with equation~(\ref{eq:muflux}) using the
GRV94 parton distributions \citep{1995ZPhyC..67..433G} and the
effective muon range from \citet{1991PhRvD..44.3543L}.\label{fig:limits}}
\end{figure}

The final results are shown in Table~\ref{Table:flux limits} and
plotted in Figure~\ref{fig:limits}, along with models of various possible
signals from AGNs \citep{Stecker:1995th,Mannheim:1998wp} and GRBs \citep{1999PhRvD..59b3002W},
as well as the backgrounds due to atmospheric neutrinos 
\citep{2004PhRvD..70d3008H,1980SvJNP..31..784V} and prompt
neutrinos \citep{2003PhRvD..67a7301G}.
%
\begin{table}
\begin{center}
\caption{Confidence intervals for the upward-going muon
flux due to neutrinos from astrophysical sources.\label{Table:flux limits}}
\begin{tabular}{cc}
\tableline 
\tableline
$E_{\mu}^{\rm min}$&
90\% C.L. range\\
(TeV)&
(${\rm {\; cm^{-2}s^{-1}sr^{-1}}}$)\\

\tableline 
$3.16$&
$0-1.03\times10^{-13}$\\
$10$&
$0-3.19\times10^{-14}$\\
$31.6$&
$0-1.26\times10^{-14}$\\
$100$&
$0-4.28\times10^{-15}$\\
\tableline
\end{tabular}
\end{center}
\end{table}

The upper limits calculated here are consistent with the models of
astrophysical signals. Also shown are models with a hypothetical isotropic
neutrino flux with a spectrum proportional to $E_{\nu}^{-2}$ and
a normalization scaled to the limits on an $E_{\nu}^{-2}$ flux set
by MACRO \citep{2003APh....19....1M} and AMANDA-II \citep{2005astro.ph..5278G}.
The model neutrino fluxes were converted into muon fluxes using equation~(\ref{eq:muflux})
as discussed in \S~\ref{sec:Analytical-Estimate}. 

To facilitate easier comparison with other experiments, we also
convert our limits on the muon flux into approximate limits on the
neutrino flux. In order to do this, we assume a model neutrino
flux that is isotropic and proportional to $E_{\nu}^{-2}$. To get
an approximate neutrino limit, we find normalization factors for an
$E_{\nu}^{-2}$ muon flux curve such that the curve passes through
each of our four limit points in Figure~\ref{fig:limits}, and we use
these factors to find the implied limits on $E_{\nu}^{-2}$ flux. 

In order to determine the approximate neutrino energy range in which
these limits are valid, we use equation~(\ref{eq:muflux}) to determine
the neutrino energy range that produces the bulk of the muon signal
for a given value of the muon energy threshold $E_{\mu}^{\rm min}$. We
define the energy range as the range that (1) produces 90\% of the
muon flux above $E_{\mu}^{\rm min}$ and (2) has a higher value of the
integrand of equation~(\ref{eq:muflux}) within the range than anywhere 
outside the range. This
definition is based on the definition of highest posterior density
intervals as described by \citet{Conway:2002}. 

\begin{figure}
\plotone{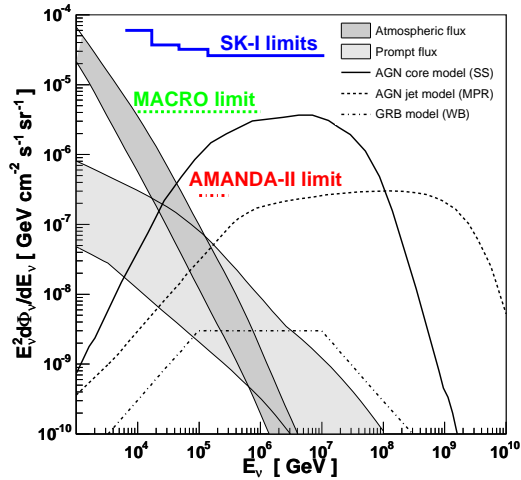}
\caption{\label{fig:neutrino-upper-limits}Approximate upper
limits from SK-I on astrophysical neutrinos $\left(\nu_{\mu}+\bar{\nu}_{\mu}\right)$.
Models shown are the same as in Figure~\ref{fig:limits}. Note that
converting from a muon flux limit to a neutrino flux limit requires
additional assumptions - our limits on the muon flux are shown in
Figure~\ref{fig:limits}. }
\end{figure}

%
%
The results of these approximations for neutrino limits and energy
ranges are shown in Table~\ref{table:neutrino-upper-limits}%
\begin{table}
\begin{center}
\caption{Approximate upper limits from SK-I on astrophysical neutrinos 
$\left(\nu_{\mu}+\bar{\nu}_{\mu}\right)$.\label{table:neutrino-upper-limits}}
\begin{tabular}{ccc}
\tableline
\tableline 
$E_{\mu}^{\rm min}$&
90\% C.L. upper limit&
Neutrino energy range\\
(TeV)&
(${\rm {GeV\, cm^{-2}s^{-1}sr^{-1}}}$)&
(GeV)\\
\tableline 
$3.16$&
$6.0\times10^{-5}$&
$6.3\times10^{3}-1.4\times10^{6}$\\
$10$&
$3.7\times10^{-5}$&
$1.7\times10^{4}-2.4\times10^{6}$\\
$31.6$&
$3.2\times10^{-5}$&
$4.7\times10^{4}-5.1\times10^{6}$\\
$100$&
$2.6\times10^{-5}$&
$1.4\times10^{5}-1.1\times10^{7}$\\
\tableline
\end{tabular}
\tablecomments{Upper limits on $E_{\nu}^{2}\left({d\Phi_{\nu}}/{dE_{\nu}}\right)$.
Note that converting from a muon flux limit to a neutrino flux limit
requires additional assumptions --- our limits on the muon flux are
shown in Table~\ref{Table:flux limits}.}
\end{center}
\end{table}
 and are also plotted in Figure~\ref{fig:neutrino-upper-limits}, %
with the same models and experimental limits as shown in Figure~\ref{fig:limits}.
To draw our limits on this plot, we chose to draw only the most sensitive
limit in each energy range.

Note that these results for the limits on the neutrino flux are only
approximations made to facilitate comparison to other experiments
--- our primary results are the limits on the muon flux shown in Table~\ref{Table:flux limits}
and Figure~\ref{fig:limits}.

\section{Conclusions}

\label{sec:Conclusions}

In conclusion, we have developed a method for analyzing Super-K's highest energy data to search for evidence of high-energy neutrino flux from astrophysical sources.  We have done a thorough study of the efficiency and the expected backgrounds from this method and applied our method to the SK-I data sample. Our study of the highest energy events in SK-I does not show evidence of a high-energy cosmic neutrino signal.

We have set upper limits on the muon flux due to cosmic neutrino sources.
These limits are consistent with the results of other experiments
\citep{2003APh....19....1M,2005astro.ph..5278G}. 
It is possible that an astrophysical neutrino signal could
be within the grasp of the next generation of neutrino detectors such
as IceCube \citep{2004APh....20..507A} and ANTARES \citep{2004EPJC...33S.971K}.

\section{Acknowledgments}
\label{sec:Acknowledgments}
We gratefully acknowledge the cooperation of the Kamioka Mining and
Smelting Company.  The Super-Kamiokande experiment has been built and
operated from funding by the Japanese Ministry of Education, Culture,
Sports, Science and Technology, the United States Department of Energy,
and the US National Science Foundation.
Some of us have been supported by funds from the Korean Research
Foundation (BK21) and the Korea Science and Engineering Foundation, 
the Polish Committee for Scientific Research (grant 1P03B08227),
Japan Society for the Promotion of Science, and
Research Corporation's Cottrell College Science Award.

\bibliographystyle{apj}
\bibliography{apj-jour,referencesaas}

\end{document}